\documentclass[aps,prb,amsmath,amssymb,reprint,superscriptaddress,preprintnumbers,showpacs,intlimits]{revtex4-1}
\usepackage{bm,mathrsfs}
\usepackage{hyperref}
\usepackage{graphicx}
\usepackage{subcaption}
\renewcommand{\vec}[1]{\bm{#1}}
\begin{document}

\title{The transmission spectra of the graphene-based  Fibonacci superlattice}%

\author{A.~M.~Korol}
\affiliation{National University for Food Technologies, Kyiv 01601, Volodymyrska str., 68, Ukraine}
\email{korolam@nuft.edu.ua}

\author{V.~M.~Isai}
\affiliation{Laboratory on Quantum Theory in Link\"{o}ping, ISIR, P.O. Box 8017, S-580, Link\"{o}ping, Sweden}

\pacs{ }

\begin{abstract}

We consider the gapped graphene superlattice (SL) constructed in
accordance with the Fibonacci rule. Quasi-periodic modulation is due to the
difference in the values of the energy gap in different SL elements. It is
shown that the effective splitting of the allowed bands and thereby forming
a series of gaps is realized under the normal incidence of electrons on the
SL as well as under oblique incidence. Energy spectra reveal periodical
character on the whole energy scale. The splitting of allowed bands is
subjected to the inflation Fibonacci rule. The gap associated with the new
Dirac point is formed in every Fibonacci generation. The location of this
gap is robust against the change in the SL period but at the same time it is
sensitive to the ratio of barrier and well widths; also it is weakly
dependent on values of the mass term in the Hamiltonian.
\end{abstract}


\maketitle

In recent years both graphene and graphene structures attracted much attention which is naturally explained by their non-trivial properties\cite{bib1,bib2,bib3,bib4,bib5,bib6,bib7,bib8,bib9,bib10,bib11,bib12,bib13}. On the other hand, notable among the semiconductor structures including graphene ones are superlattices intermediate between periodic and disordered -- quasi-periodic structures, e.g. Fibonacci, Thue-Morse and others. This is due to their unusual properties, such as self-similarity, fractal-like electronic spectrum etc. Attempts has already been started to study some kinds of superlattices based on graphene (Fibonacci and Thue-Morse \cite{bib10,bib11}) which shed light on the behaviour of Dirac chiral fermions in the quasi-periodic chains. Motivated by these circumstances, in this report we study the energetical spectra of the Fibonacci SL based on the monolayer gapped graphene. Quasi-periodic modulation is due to the difference in the values of the mass term $\Delta $ in the Hamiltonian in different elements of the superlattice. Because of the substantial progress in techonology of graphene structures, in particular concerning the fabrication of the gapped graphene \cite{bib14,bib15,bib16,bib17,bib18,bib19,bib20} we vary the value of $\Delta $ in wide range regardless of its nature which may be different \cite{bib14,bib15,bib16,bib17,bib18,bib19,bib20}.

\begin{figure}
\includegraphics[width=\columnwidth]{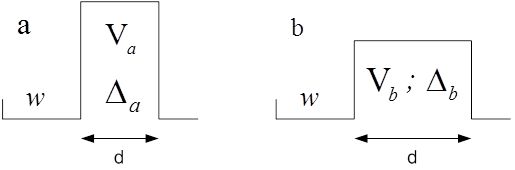}
\caption{}
\label{fig0}
\end{figure}

Consider the SL built of two elements ``a'' and ``b'', see Fig.~\ref{fig0}.

Both elements contain the quantum well of width ``w'' and a potential
barrier of width ``d''. The term $\Delta _{a}$ corresponds to the ``a''
element and the barrier height is denoted as $V_{a}$, for an element ``b''
we have $\Delta _{b}$ and $V_{b}$ respectively. The SL is constructed
according to the Fibonacci inflation rule. Thus for the fifth Fibonacci
generation one can write : $S_{5}$=abaababa. The value of the transmission
coefficient $T(E)$, $E$ being the electron energy, through the SL built for a
certain generation is determined by the period of this generation
(sequence). Energy intervals in which the condition $T(E)=1$ holds form the allowed bands in energy spectrum, gaps are associated with values $T\ll 1$. The transmission coefficient can be evaluated by the
transfer matrix method, by expressing $T$ either in terms of the Green
functions or the eigenfunctions. The latter are the solutions of the
Dirac-like equation
\begin{equation} \label{eq1}
\left[ {v_F (\vec {\sigma },\vec {p})+m^2\cdot v_F ^2\cdot \sigma _z +V(x)\cdot \hat {I}} \right]\cdot \Psi =E\cdot \Psi,
\end{equation}
where $v_{F}=10^{6} m/s$, $\vec{p}=(p_{x},p_{y})$ the momentum operator, $\sigma = (\sigma_x,\sigma_y)$, $\sigma_{x}$, $\sigma_{y}$, $\sigma_{z}$ --- Pauli matrices for the pseudospin, $V(x)$ the external potential, which depends only on coordinate $x$, $I$ -- two-dimensional unit matrix, the mass term is denoted by the symbol $\Delta $ as adopted in the literature. The function $\Psi $ is a two-component pseudospinor $\Psi =[\Psi_{A}, \Psi_{B}]^{T}$, $\Psi_{A}$, $\Psi_{B}$ are the envelope functions for the graphene sublattices $A$ and $B$. Suppose that the potential consists of repetitive rectangular barriers along the $x$-axis and $V_{j}(x)= const$ within each $j$-th barrier. Since the $y$-component of the electron momentum commutes with the Hamiltonian one can write $\tilde{\Psi }_{A,B} =\Psi_{A,B} \cdot e^{i k_y \cdot y}$ and we get from (\ref{eq1}):
\begin{equation} \label{eq2}
\frac{d^2\Psi _{AB} }{dx^2} +\left( k_j^2- k_y^2 \right)\Psi _{A,B} =0
\end{equation}
where $k_{j} =\text{sign}(s_{j+} )[(E-V_j)^2-\Delta^2]^{1/2}$, $s_\pm \equiv E-V(x)\pm \Delta $, where units $c=h=e=v_{f}=1$ are adopted. If we present solutions for the eigenfunctions $\Psi_{A,B}$ as a sum of plane waves traveling in the forward and backward directions of the $x$-axis then we obtain:
\begin{equation} \label{eq3}
\Psi (x)=\left[ a_{j} \cdot e^{i\cdot q_j \cdot x}
\begin{pmatrix}
1\\
g_j^+
\end{pmatrix}
+b_j \cdot e^{-i\cdot q_j \cdot x}
\begin{pmatrix}
1\\
g_j^-
\end{pmatrix}
\right]
\end{equation}
where $q_{j} =\text{sign}(s_{j^+} )\sqrt {k_j^2 -k_y^2 } $ if $k_{j}^{2}>k_{y}^{2}$ and $q_j =i\cdot \sqrt {k_y^2 -k_j^2 }$ otherwise, ${g}_j ^\pm =\left({\pm q_j +i\cdot k_y }\right)/{k_y },$ the top line in (\ref{eq3}) pertains to the sublattice $A$, the lower one --- to the sublattice $B$. Transfer matrix connecting the wave the functions at the points $x$ and $x + \Delta x$ is found in several studies (see e.g.\cite{bib6}) and has the form:
\begin{equation} \label{eq4}
{M}_{j} =\frac{1}{\cos \theta _j }
\begin{pmatrix}
{\cos (q_j \cdot \Delta x-\theta _j )} & {i z_j^{-1} \sin (q_j\cdot \Delta x)} \\
{i z_j \sin (q_j \cdot \Delta x)}      & {\cos (q_j \cdot \Delta x+\theta _j )}
\end{pmatrix}
\end{equation}
where ${z}_{j} =\frac{s_{j^-} }{k_j }$ , $\theta_j =\arcsin \left({\frac{k_y }{k_j }} \right)$.

\begin{figure}
\includegraphics[height=\columnwidth]{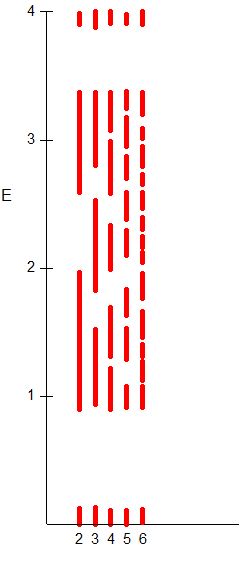}
\caption{Trace map for the Fibonacci SL, $k_{y }=0$, $d=w=0,5$, $V_{a}=V_{b}=1$, $\Delta _{a}=1$, $\Delta _{b}=0$.}
\label{fig1}
\end{figure}

The coefficient of the transmission of quasi-particles through the lattice $T=|t|^{2}$,
\begin{equation} \label{eq5}
t=\frac{2\cos \theta _0 }{R_{22} e^{-i\cdot \theta _0 }+R_{11} e^{i\cdot \theta _0 }-R_{12} -R_{21} },
\end{equation}
where $\theta_{0}$ is the angle of incidence, and the matrix $R$ is expressed as a product of matrices $M_{j}$: $R=\prod\limits_{j=1}^N {M_j }$, $n$ --- the number of elements in the SL.

The trace map for 6 initial Fibonacci sequences is depicted in Fig.~\ref{fig1} for the
case where we have the gapped graphene in elements ``a'' and the gapless one
in elements ``b''. First of all it is noteworthy that the quasi-periodic
modulation by setting different values of $\Delta $ in different SL elements
used in this study results in the efficient splitting of the allowed energy
bands and thus in the formation of a number of gaps. And this is realized at
normal incidence of the electrons on the lattice.

\begin{figure}
\includegraphics[width=\columnwidth]{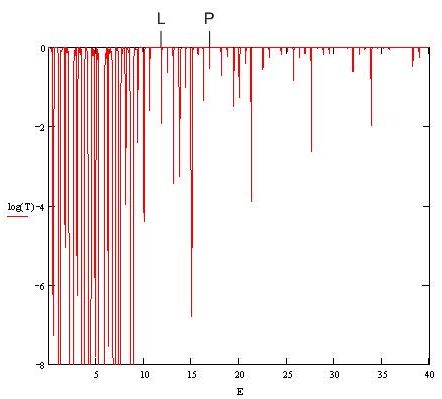}
\caption{Dependence of the transmission coefficient $T$ on electrons energy $E$ for the 4-th Fibonacci generation; values of the parameters: $k_{y }=0$, $d=w=0,5$, $V_{a}=V_{b}=5$, $\Delta_{b}=0$.}
\label{fig2}
\end{figure}

\begin{figure}
\includegraphics[width=\columnwidth]{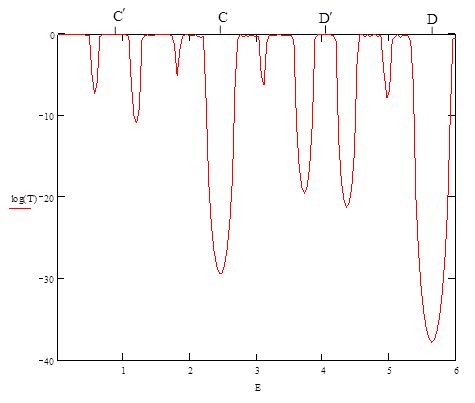}
\caption{Tunneling spectrum for the 4-th Fibonacci generation in the energy range $[0,6]$; values of the parameters are the same as those in Fig.~\ref{fig2}.}
\label{fig3}
\end{figure}

Fig.~\ref{fig2} exhibits the tunneling spectrum, i.e. the dependence of the
transmission coefficient T on energy E for the 4-th Fibonacci generation
under normal incidence of the electron wave on the lattice; the same
spectrum in the energy range $[0, 6]$ is shown in Fig.~\ref{fig3}. We see that the
structure of some parts of the spectra is repeated periodically throughout
the whole energy scale; one of these fragments of the spectrum can be
considered as its period. The characteristic features of the period -- the
number of allowed (forbidden) bands and their widths change so that with E
increasing the gap's widths on average decrease. The natural result of this
reduction is that the transmission coefficient asymptotically approaches to
unity in the sufficiently far over-barrier region.

The band widths don't decrease monotonically -- with E increasing the
alternate extension and reduction of gaps (or the allowed bands) is
observed. This results in a wider periods (like a super-period) as it is
clearly seen in Fig.~\ref{fig1} (e.g. the interval between the points $L$ and $P$). In
other words, the result of such a wavelike changes of band widths is the
grouping of smaller structural units into bigger ones to form the additional
structural order -- a manifestation of the self - similarity in the problem
considered.

Spectra similar to that shown in Fig.~\ref{fig2} for the 4-th generation are observed
also for other Fibonacci sequences.

The number of bands and the width of each of them depend on one hand on the
SL parameters and on the other hand on the number of the generation.

Here we would like to draw attention to a certain contrast to the situation
in conventional SL (with a parabolic dispersion law of the charge carriers).
Unlike in the conventional SL where the calculation of bands usually is
carried out within the barrier region, in graphene Fibonacci structures for
this purpose it is advisable to choose the certain energy intervals, e.g.
periods of spectra (see Figs.~\ref{fig2}, \ref{fig3}) or some other fixed fragments of spectra.

As follows from the calculations, the number of allowed (forbidden) bands
contained in one period, in particular in the interval CD in Fig.~\ref{fig3},
corresponds to the Fibonacci sequence and this number is subjected to the
Fibonacci inflation rule: $Z_{n}=Z_{n-1 }+Z_{n-2}$, where $n$ is the
number of the Fibonacci generation (see Fig.~\ref{fig1}). Note that this rule is
applied not only to the CD period but to larger periods as well; every new
super-period has its own number of bands.

\begin{figure}
\includegraphics[width=\columnwidth]{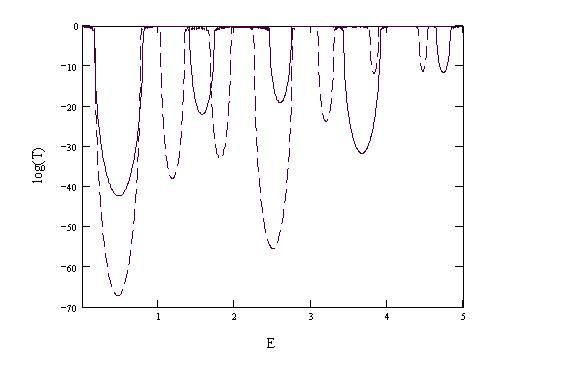}
\caption{}
\label{fig4}
\end{figure}

As we can see in Figs.~\ref{fig2}, \ref{fig3} at a certain energy all Fibonacci generations form the gap associated with the new Dirac point -- ``new Dirac point gap'' \cite{bib6}. Its location is almost unchanged in different Fibonacci sequences. A typical feature of the new Dirac point gap is that it is independent of the lattice period $(d + w)$ and at the same time is sensitive to the ratio $d/w$. This is proved in particular by Fig.~\ref{fig4} which shows the dependence of $T$ on $E$ for the third Fibonacci generation for different values of $w$ and $d$. Once again we draw attention to the fact that the band's splitting (thereby forming a series of gaps) is realized even in the case of the normal incidence of electrons on the lattice surface. This result is significantly different from what was obtained in \cite{bib10} in which a quasi - periodic modulation was created by the difference in the potentials of elements ``a'' --- the barrier and ``b'' --- the well, and the splitting of bands was observed when $k_{y }\neq 0$ only (oblique incidence of the wave).

The location of the new Dirac point depends in general on values of each parameter $d$, $w$, $V_{a }$, $V_{b }$, $\Delta _{a}$ , $\Delta _{b}$ and it turns out that if we put $V_{a} =V_{b}=V$, $d=w$ then this position is equal to $E_{d}\approx \frac{V}{2}$ and only slightly deviates from this value with increasing of $\Delta({V}/{2})$ is the exact position of the new Dirac point in the periodic lattice \cite{bib6}).

\begin{figure}
\includegraphics[width=\columnwidth]{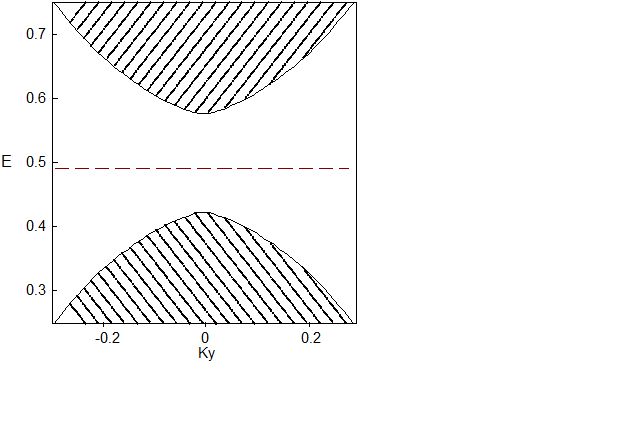}
\caption{}
\label{fig5}
\end{figure}

The band structure for the SL of the third Fibonacci generation is shown in Fig.~\ref{fig5} in coordinates $E$, $k_{y}$; it features the spectra dependence on the angle of incidence $\theta $. Note that there is some extension of the new Dirac gap and at the same time the dependence of other gaps on $\theta $ is very weak (not shown in Fig.~\ref{fig5}). This circumstance is known to be common in the study of certain effects in graphene structures, in particular, the same result was stated in \cite{bib8}: if there is a sufficiently strong effect at $k_{y}$ = 0 then the dependence on $k_{y}$ is weak, see also the comment in \cite{bib8}. The dashed line in Fig.~\ref{fig5} denotes the new Dirac point's location (almost equal to $V/2$).


\begin{thebibliography}{20}%

\bibitem{bib1} %
A.K.Geim, K.S.Novoselov, Nat. Materials \textbf{6},183 (2007).

\bibitem{bib2} %
A.N.Castro Neto, F.Guinea, N.M.R.Peres et al., Rev. Mod. Phys. \textbf{81},109 (2009).

\bibitem{bib3} %
J.M.Pereira, F.M.Peeters, A.Chaves et al., Semicond. Science Technology \textbf{25},033002 (2010).

\bibitem{bib4} %
V.V.Cheianov, V.I. Falko, Phys. Rev. B \textbf{74}, 041403 (2006).

\bibitem{bib5} %
Q.Zhao, J.Gong, C.A.Muller, Phys. Rev. B \textbf{85}, 104201 (2012).

\bibitem{bib6} %
L.Wang, X.Chen, J.Appl. Phys. \textbf{109}, 033710 (2010).

\bibitem{bib7} %
L.Wang, S.Zhu, Phys. Rev.B \textbf{81}, 205444 (2010).

\bibitem{bib8} %
V.H.Nguyen, A.Bournel, P.Dollfus, Semicond.Sci.Technol. \textbf{26},125012 (2011).

\bibitem{bib9} %
M.Barbier, P.Vasilopoulos, F.M.Peeters, Phys. Rev.B \textbf{80},205415 (2009).

\bibitem{bib10} %
P.Zhao, X.Chen, Appl.Phys. lett. \textbf{99}, 182108 (2011).

\bibitem{bib11} %
T.Ma, C.Liang,L.Wang et al., Appl.Phys. lett. \textbf{100}, 252402 (2012).

\bibitem{bib12} %
Yu.P.Bliokh, V.Freilikher, S.Savel'ev et al., Phys. Rev. B \textbf{79}, 075123 (2009).

\bibitem{bib13} %
P.V.Ratnikov, JETP Letters \textbf{90}, 469 (2009).

\bibitem{bib14} %
J.C.Meyer, C.O.Girit, M.F.Crommie et al., Appl. Phys. Lett. \textbf{92},123110 (2008).

\bibitem{bib15} %
P.W.Sutter, J.Flege, E.A.Sutter, Nat. Materials. \textbf{7}, 406 (2008).

\bibitem{bib16} %
J.Coraux, A.T.N'Diaye, C.Busse et al., Nano Lett.\textbf{ 8},565 (2008).

\bibitem{bib17} %
Y.W.Son, M.L.Cohen, S.G.Louie, Phys.Rev.Lett \textbf{97},216803 (2006).

\bibitem{bib18} %
M.Y.Han, B.Ozyilmaz, Y.Zhang et al., Phys. Rev. Lett. \textbf{98}, 206805 (2007).

\bibitem{bib19} %
G.Giovanetti, P.A.Khomyakov, G.Brocks et al., Phys. Rev. B \textbf{76}, 073103 (2007).

\bibitem{bib20} %
S.Y.Zhou, G.Gweon, A.V.Fedorov et al., Nat. Materials \textbf{6},770 (2007).

\end{thebibliography}
\end{document}